\RequirePackage{fix-cm}
\documentclass[smallextended]{svjour3}       
\smartqed  
\usepackage{graphicx}
\usepackage{epstopdf}
\usepackage{geometry}
\usepackage{graphicx,amssymb,amstext,amsmath}
\usepackage[justification=centering]{caption}

\usepackage{epsfig}
\usepackage{subfigure}
\usepackage{amsmath}
\usepackage{algorithmic}
\usepackage{graphics}
\usepackage{amssymb}
\usepackage{alltt}
\usepackage{float}
\usepackage{array}
\usepackage{xspace}
\usepackage{footmisc}
\usepackage{comment}
\usepackage{algorithm}
\usepackage{algorithmic}
\usepackage[misc]{ifsym}

\usepackage{multirow}
\usepackage{colortbl}

\begin{document}

\title{Temporal Activity Path Based Character Correction in Social Networks
}
\subtitle{}


\author{Jun Long$^{\dagger \S}$         \and
        Lei Zhu$^{\dagger \S}$         \and
        Zhan Yang$^{\dagger \S}$       \and
        Chengyuan Zhang$^{\dagger \S}$  \and
        Xinpan Yuan$^{\ddagger}$      \and
}


\institute{Jun Long \at
              \email{jlong@csu.edu.cn}           
           \and
           Lei Zhu \at
              \email{leizhu@csu.edu.cn}
           \and
           Zhan Yang \at
              \email{zyang22@csu.edu.cn}
           \and
           \Letter Chengyuan Zhang \at
              \email{cyzhang@csu.edu.cn}
           \and
           Xinpan Yuan \at
              \email{xinpanyuan@outlook.com}
           \and
           \\
           \\
           $\dagger$ School of Information Science and Engineering, Central South University, PR China \\
           $\S$ Big Data and Knowledge Engineering Institute, Central South University, PR China \\
           $\ddagger$ School of Computer, Hunan University of Technology, PR China
}

\date{Received: date / Accepted: date}

\maketitle

\begin{abstract}
Vast amount of multimedia data contains massive and multifarious social information which is used to construct large-scale social networks. In a complex social network, a character should be ideally denoted by one and only one vertex. However, it is pervasive that a character is denoted by two or more vertices with different names, thus it is usually considered as multiple, different characters. This problem causes incorrectness of results in network analysis and mining. The factual challenge is that character uniqueness is hard to correctly confirm due to lots of complicated factors, e.g. name changing and anonymization, leading to character duplication. Early, limited research has shown that previous methods depended overly upon supplementary attribute information from databases. In this paper, we propose a novel method to merge the character vertices which refer to as the same entity but are denoted with different names. With this method, we firstly build the relationship network among characters based on records of social activities participated, which are extracted from multimedia sources. Then define temporal activity paths (TAPs) for each character over time. After that, we measure similarity of the TAPs for any two characters. If the similarity is high enough, the two vertices should be considered to the same character. Based on TAPs, we can determine whether to merge the two character vertices. Our experiments shown that this solution can accurately confirm character uniqueness in large-scale social network.
\end{abstract}

\section{Introduction}
\label{Intro}

In the past decade, the mobile Internet and social multimedia applications have become an indispensable part of social life, and huge multimedia data are being produced and consumed~\cite{DBLP:conf/ism/SangXJ16}. For instance, Facebook reports 350 million photos uploaded daily as of November 2013; 100 hours of video are uploaded to YouTube every minute, resulting in more than 2 billion videos totally by the end of 2013~\cite{DBLP:conf/bigmm/SangX15}. Social Media Networks allow people to communicate, share, comment and observe different types of multimedia content~\cite{DBLP:SOSAmultimedia}. As social activities are becoming more frequent, social networks have been larger and much more complex. Generally, we extract information and construct social transaction databases from vast amount of multimedia data, such as text, images~\cite{DBLP:journals/tip/WangLWZ17,DBLP:journals/mta/WuHZSW15,DBLP:conf/mm/WangLWZ15}, videos and audios~\cite{DBLP:journals/jvcir/NiePWZS17,DBLP:conf/sigir/WangLWZZ15,DBLP:conf/cikm/WangLZ13}, to construct large-scale social networks which are modelled by graphs with node-edge representation. Multimedia data, generally, can be described in multi-views~\cite{NNLS2018,DBLP:conf/ijcai/WangZWLFP16} such as color view and textual view~\cite{DBLP:journals/tnn/WangZWLZ17,DBLP:journals/tip/WangLWZZH15,DBLP:conf/pakdd/WangLZW14}. In social networks, the relations between character vertices are tangled by the time difference of transaction, incompletion of personal information record, anonymous phenomena and difference of information pattern and structure. It is distinctly difficult to maintain one-to-one mapping between characters in relation networks and people in real life. Besides, characters are marked up as difference vertices by former and present name. These vertices have the same personal information, structure and attributes of relation. For social networks, these vertices and relationships are redundant, which will severally perturb the results of social network analysis. Therefore, character vertices ambiguity has become a key problem in social network analysis.

In relational databases, we can use multidimensional personal information to confirm uniqueness of characters, such as name, gender, date of birth, etc. In big data environment, however, multimedia data is mainly from unstructured data storage. Its scale is vast, types are multifarious~\cite{DBLP:conf/cts/SagirogluS13,DBLP:conf/ic3/KatalWG13,DBLP:journals/corr/abs-1708-02288}, such as text, images, videos, audios etc. Besides, the data is not complete and consistent generally, which caused the uniqueness of characters to be difficult to determine. Most of the large-scale multimedia data, nevertheless, are stored externally, and included people's participation in social activities, such as vocational and educational experience, participate in service club etc. Social relationships are generated by using multimedia data of these activities, and then the social networks can be built up. It is a key problem to confirm the character uniqueness in social networks analysis and application. We propose to measure the similarity of characters by network characteristics to conduct character correction by vertices merging with computing structure error of networks. However, social networks have temporal attributes generally, and relations extracted from them also have obvious temporal characteristics. We regard temporal attributes as a key factor of relations, which is used to computing the similarity of vertices. Accordingly, it boosts the accuracy of uniqueness conforming. We put forward new notions of character activity path and transaction activity network with heterogeneous features, and then use temporal activity path similarity evaluation to improve the reliability of character correction.

The remainder of this paper is organized as follows: We introduce related work in section 2, and then describe an academic network building process in section 3. In section 4, we present character vertices merging principle and structure error algorithm based on network structure. In section 5, we introduce transaction activity networks, activity paths algorithm and vertices screening, and then analyze the experiment results. The conclusion and future work are included in section 6.

\section{Related work}
\label{related}
In this section, we present an overview of existing researches of image retrieval and ranking, spatial keyword query and interactive search, which are related to our study. To the best of our knowledge, there is no existing work on the problem of interactive geo-tagged image search.

\subsection{Social Analysis via Multimedia}
The advent of social networks and cloud computing has made social multimedia sharing in social networks become easier and more efficient~\cite{DBLP:journals/vlc/YeXDWLZ14,DBLP:journals/ivc/WuW17}. With the rapid increasing of volume of multimedia data, social networks analysis and mining via multimedia data attract attention of a number of researchers recently. Zhuhadar et al. proposed combination of social learning network analysis and social learning content analysis in studying the impact of the social multimedia systems cyberlearners~\cite{DBLP:journals/chb/ZhuhadarYL13,DBLP:journals/pr/WuWGL18}. They presented evidence obtained from the analysis that Social Multimedia System impacts the communication between faculty and students. To deal with the challenges of event detection rom massive social media data in social networks, Zhao et al.~\cite{DBLP:journals/zhaosgaoy,DBLP:journals/cviu/WuWGHL18,YXWINF13,YXJPAKDD14,LYJMM13,LYSARX} proposed a novel real-time event detection method named microblog clique to explore the high correlations among different microblogs, which was supported by social multimedia data. Sang and Xu~\cite{DBLP:conf/bigmm/SangX15,TC2018} proposed to analyze into variety of big social multimedia from the perspective of various sources. Laforest et al.~\cite{DBLP:conf/mownet/LaforestSFCMLGS14,DBLP:journals/pr/WuWLG18} present a new kind of social networks named spontaneous and ephemeral social networks (SESNs) which allow people to collaborate spontaneously in the production of multimedia documents. In order to finding overlapping communities from multimedia social networks, Huang et al.~\cite{DBLP:journals/tmm/HuangLZZCZ17} proposed an efficient algorithm named LEPSO for overlapping communities discovery, which is based on line graph theory, ensemble learning, and particle swarm optimization.

\subsection{Name Ambiguity and De-anonymity}
Recently, name ambiguity and de-anonymity have been widely studied. There are several methods to identify characters in social networks, which can be divided into three categories, for   internal relational database, Internet webpage and topology structure of social networks.

Name ambiguity and de-anonymity are with the same essential features. In the past, the identity of the characters is determined by the accurate attribute information in internal databases of enterprises. In 2008 Arvind Narayannan and Vitaly Shmatikov proposed the method to process high dimensional data~\cite{DBLP:conf/sp/NarayananS08}, such as personal attribute, recommend and transaction information etc. Users can identify characters in the anonymous database with limited personal information. This method has strong robust even though background information is inaccurate or disturbed. However, internal database is localized and static, it cannot describe feature of characters thoroughly. Therefor these methods are not suitable for name ambiguity problem in big data with complexity, dynamicity and cross platform.

Name ambiguity is more prominent in the Internet. In 2008 Jie Tang and Jing Zhang et al. proposed a standard probability framework to recognize the independence of observed objects~\cite{DBLP:conf/kdd/TangZYLZS08}. But when we search name from the Internet, numerous webpages containing one same name can be returned, and it is not sure that these pages are belong to one same people. Ron Bekkerman and Andrew McCallum proposed two statistical methods to solve this problem in 2005~\cite{DBLP:conf/www/BekkermanM05}. The one is based on link structure of webpages and the other is on multi-way distributional clustering method, which is unsupervised frameworks and only need a few of prior knowledge, and the experiments show that their solution outperform traditional clustering. However, above methods are deeply subject to the uncertainty of web information.

At present, name disambiguation becomes even more prominent in social networks modeling and analysis. In 2008, Kun Liu and Evimaria Terzi analyzed character-centered social networks and then pointed out that features of relation structure can expose characters’ identity~\cite{DBLP:conf/sigmod/LiuT08}. For identity hiding in social relations network, they defined graph-anonymization and proposed the algorithm based on k-degree anonymous graph and node degree sequence. Arvind Narayanan and Vitaly Shmatikove defined privacy of social networks in 2009~\cite{DBLP:conf/sp/NarayananS09}, and designed a novel re-identification algorithm which can implement de-anonymity and identifying node by using the topological structure of network. For dynamic evolution of social networks, Xuan Ding and Lan Zhang et al. proposed the “threading” technique and used the connection between released data to implement de-anonymizing~\cite{DBLP:conf/globecom/DingZWG11}. And they proposed to combine structure information and attributes of nodes to re-identify anonymous nodes. Mohammed Korayem and David J. Crandall worked on de-anonymizing method specially in cross platform social networks~\cite{DBLP:conf/icwsm/KorayemC13}, which can recognize different accounts belong to one user by extracting time sequence data, text features, geographic location and social relation characteristics. In 2012, Mudhakar Srivatsa and Mike Hicks introduced de-anonymizing users’ mobile trace information based on graph structure of social networks~\cite{DBLP:conf/ccs/SrivatsaH12}. As contact graph between characters consist of vast quantities of mobile trace, they proposed structure similarity of inter-user correlation, which was used to map in contact graph and social network.

Since a large number of mobile trajectory can be used to build the contact graph of characters, the structural similarity is employed to find out the corresponding nodes in the contact graph and social network. The de-anonymity with mobile trace is implemented by mapping character nodes between the two networks. Above methods, it is the anonymous in traditional networks, nodes and relations have only one category respectively. However, the social network created on big data is comprehensive, such as category diversity of relationship and nodes, temporality etc. In addition, these methods need to add attributes to supplement topology information of social networks.

From enterprise databases to webpages databases and then to social networks databases, this is a developing process from local data to global data. Previous methods rely on attribute details of local data, namely it needs much more auxiliary information to identify characters, but the efficiency is low. However, big data is multi-sourced~\cite{DBLP:journals/tkde/WuZW014}, time-variable, global and macroscopic. The social networks are built on global data, and it is impossible to look back upon distributed sources. In this type of networks, intrinsic relationship structure of vertices is a key factor to measure uniqueness of characters. Since different characters have different social relationships, we can identify characters by network structure features. For the heterogeneity and temporality of big data networks, we propose uniqueness correction method and the notion of activity path similarity based on heterogeneous temporal networks~\cite{DBLP:journals/pvldb/SunHYYW11}, to promote the efficiency and accuracy of character identification.

\section{Social Network Modeling}
\label{SocialNetworkModeling}

A large-scale social network is based on diversified multimedia data which is multi-modality~\cite{DBLP:conf/mm/WangLWZZ14}, for instance, an image can be described by color modality or shape modality. It contains information of multifarious and complex social activities. We can build this kind of network by extracting relations from transaction activity information which is extracted from multimedia datasets. As academic network is a typical case of social network, we use it as an example to describe the process of social networks mining.

In general, academic relations mainly include teacher-student relationship, classmate relationship, project partnership, and co-author relationship etc. These relations are contained in education experiences, research and work experiences, cooperation and co-author experiences, and academic activities and conferences experiences etc. The information of academic activities is contained in project proposals, project progress and concluding reports, degree certificates, award certificates, photographs or videos concerning conference and other scientific information documents. Therefore, we extract academic activities information from the multimedia data and then construct academic transaction activity network. It is the base of mining and analysis of academic network between scholars.

\begin{figure}[thb]
\newskip\subfigtoppskip \subfigtopskip = -0.1cm
\centering
\includegraphics[width=0.7\linewidth]{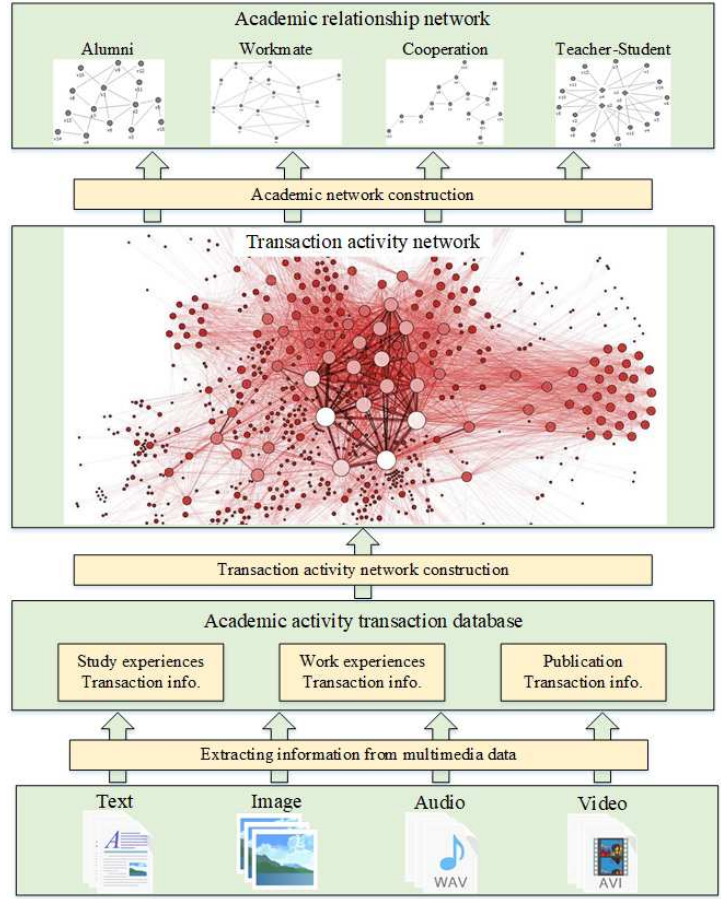}
\vspace{-1mm}
\caption{\small  Framework of academic relationship network construction }
\label{fig:fig1}
\end{figure}

Figure~\ref{fig:fig1} shows a general view of framework of academic relation network construction. First, using academic activity transaction extract method from multimedia sources which contain texts, images, audios and videos to collect individual resume information of scholar, team members information etc. Then we construct an academic activity transaction database. This database contains personal information, study and work experience information, project and publication information etc. After that, we build academic activity relation network contained heterogeneous vertices and relations. On this basis, we create academic transaction activity networks. In this kind of networks, there are several types of transaction activities, such as study experiences (‘graduated from Tsinghua University’, ‘studying in Cambridge University’, ‘was conferred doctor’s degree’, etc.), work experiences (‘worked at Microsoft MSRA’, ‘teaches in Central South University’, etc.), publication information (‘published‘ ($\mu$ + $\lambda$) Evolutionary Strategy for 3D Modeling and Segmentation with Super quadrics’, etc.), research experiences (‘took over The Association Rules Mining of Time Series and Knowledge Discovery for Recognition of Expert Academic Activities Track project’, etc.), etc.

These transaction networks are 2-mode networks which consist of two types of vertices: character vertices and entity vertices, and their activity relations. These vertices represent scholars or researchers, and academic entities respectively. We can mine alumni relationship, workmate relationship, project cooperation and co-author relationship from them. The character vertices and academic relation constitute academic relationship network which is a kind of homogeneous 1-mode network. We proposed vertices merging method based on structure error of network to implement uniqueness correction in this 1-mode network.

\section{Evaluating Uniqueness of Character Vertices Based on Structure Error}
\label{EvaluatingUniqueness}

Redundant information of vertices and relation is generally carried by non-unique character vertices. Thus, correct structure merging is a key process to remove redundant information from social networks. Theoretically, structure of networks will not be changed after redundant vertices and relations merging. We evaluated uniqueness of character vertices by merging test and then screened out redundant vertices candidates.

\subsection{Evaluating Uniqueness of Character Vertices}
In a social network, we consider the character vertices which have the same neighbor as suspicious redundant vertices. Some of them containing redundant information are non-unique, and the others with a high similarity may not be redundant. Thus, we call suspicious redundant vertices as redundant vertices candidates.

\subsubsection{Uniqueness of vertices}
Let $G=<V,R,\Phi>$ be a 1-mode network in which the vertices represent characters. Two nonempty finite sets $V$ and $R$ are character vertices set and relations set. We denote the mapping from relations set to vertices set as $\Phi: R \rightarrow {<v_x,v_y>|v_x \in V \and v_y \in V}$. The set which has $I$ vertices to be tested is denoted as $V_I$ and $\hat{V}_J$ is set of $J$ neighbor vertices, and $V_I
\bigcap \hat{V}_J = \emptyset, \{v_1,v_2,...,v_I\} \subseteq V_I$, $\{\hat{v}_1,\hat{v}_2,...,\hat{v}_J\} \subseteq \hat{V}_J$. The relation set which contain $K$ relations between character vertices and neighbor vertices is denoted as $R=\{r_1,r_2,...r_K\}$, and $|R|=K$. The mapping from character vertices set to relations set is denoted as $\psi:V_I \rightarrow R$, and the mapping from character vertices to its neighbor vertices is denoted as $\Lambda:V_I \rightarrow \hat{V}_J$.

\textbf{Property 1}: In a 1-mode network $G$, if vertices $v_1,v_2,...v_I$ in the vertices set $V$ have uniqueness, then $\Lambda(v_I) \neq \Lambda(v_2) \neq...\neq \Lambda(v_I)$ and the values of structure error between $V$ are not zero.

\subsubsection{Redundant Vertices Candidates}
In theory, character vertices which are non-unique have selfsame or nearly identically relation structure. Redundant relations and vertices are generated by this situation and they should be merged so as to remove redundant information. We introduce the notion of structure error to describe the difference of network structure between vertices. The vertices with selfsame or highly similar structure are referred to as redundant vertices candidates. They contain redundant relation information.

\begin{definition}[\textbf{Redundant vertices candidates}] \label{def:igis rvc}
In a network $G=<V,R,\Phi>$, character vertices $v_1,v_2,...v_I$  are redundant vertices candidates if the values of structure error between them are zero, and the redundant vertices candidate set is denoted as $H=\{v_1,v_2,...,v_I\}$.
\end{definition}

\begin{definition}[\textbf{Redundant vertices}] \label{def:igis rv}
Let a vertices set be $\widetilde{H}=\{\widetilde{H}_1,\widetilde{H}_2,...\widetilde{H}_n\}$, $\widetilde{H}_1=\{v_1,v_2,...v_{\mu_1}\}$, $\widehat{H}_2=\{v_{\mu_1+1},v_{\mu_1+2},...,v_{\mu_2}\}$,$...$,$\widetilde{H}_n=\{v_{\mu_{n-1}+1},v_{\mu_{n-1}+2},...,v_{\mu_n}\}$.
If the vertices in $\widetilde{H}_1,\widetilde{H}_2,...,\widetilde{H}_n$ are non-unique, the set $\widetilde{H}$ is referred to as redundant vertices set. The number of all vertices in $\widetilde{H}$ is denoted as $\mu_n$.
\end{definition}

If $\widetilde{H}=\{\widetilde{H}_1,\widetilde{H}_2,...,\widetilde{H}_n\}$ is the redundant vertices set, the vertices merging process is: $\forall v_i \in \widetilde{H}_1,v_{i+1} \in \widetilde{H}_2,...,v_{i+n} \in \widetilde{H}_n$, and calculate $N_1=\widetilde{H}_1-\{v_i\}$,$N_2=\widetilde{H}_2-\{v_{i+1}\}$,...,$N_n=\widetilde{H}_n-\{v_{i+n}\}$ and $E_1=\{r_1,r_2,...,r_K\}-\Psi(v_i)$,$E_2=\{r_1,r_2,...,r_K\}-\Psi(v_{i+1})$,...,$E_n=\{r_1,r_2,...,r_K\}-\Psi(v_{i+n})$. Then calculate $V'=V-(N_1 \bigcup N_2 \bigcup...\bigcup N_n)$ and $R'=R-(E_1 \bigcup E_2 \bigcup...\bigcup E_n)$  to get the merged network $G'=<V',R',\Phi>$. This new network does not contain any redundant information because $\mu_n-n$ vertices have been removed by vertices merging.

\begin{figure}[thb]
\newskip\subfigtoppskip \subfigtopskip = -0.1cm
\centering
\includegraphics[width=1\linewidth]{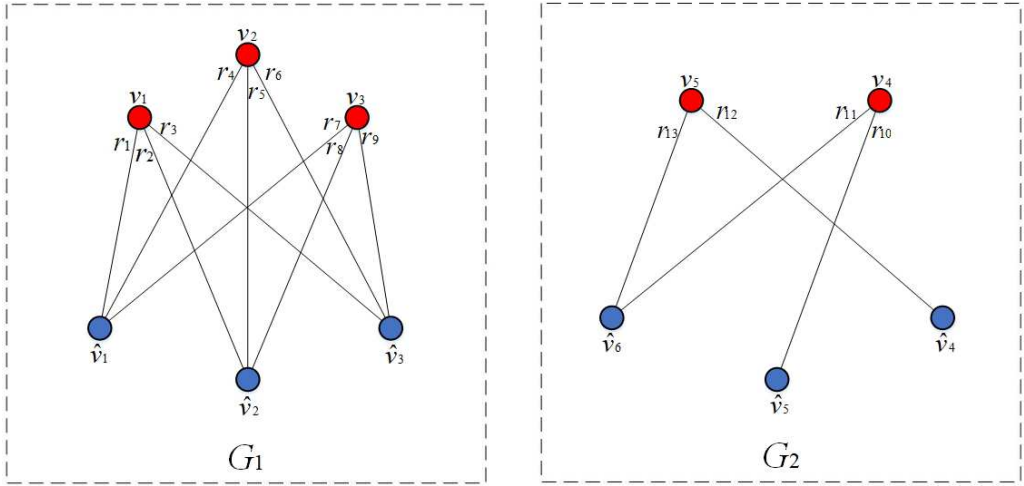}
\vspace{-1mm}
\caption{\small  Redundant vertices candidate construction }
\label{fig:fig2}
\end{figure}

In Figure~\ref{fig:fig2}, network $G_1$ contains six character vertices and nine relations, we denote them as $V=\{v_1,v_2,v_3,\hat{v}_1,\hat{v}_2,\hat{v}_3\}$ and $R=\{r_1,r_2,r_3,r_4,r_5,r_6,r_7,r_8,r_9\}$, the neighbors of vertices $v_1$, $v_2$ and $v_3$ are denoted as $\hat{v}_1,\hat{v}_2,\hat{v}_3$. Before merging process, they connect with neighbor vertices $\hat{v}_1,\hat{v}_2,\hat{v}_3$ respectively. Therefore $\Lambda(v_1)=\Lambda(v_2)=\Lambda(v_3)=\{\hat{v}_1,\hat{v}_2,\hat{v}_3\}$,
we regard $v_1$, $v_2$ and $v_3$ as redundant vertices candidates. In network $G_2$, vertices set and relation set are $V=\{v_4,v_5,\hat{v}_4,\hat{v}_5,\hat{v}_6\}$. Both $v_4$ and $v_5$ have two relations but structure of them is different, namely $\Lambda(v_4)=\{\hat{v}_5,\hat{v}_6\}$,$\Lambda(v_5)=\{\hat{v}_4,\hat{v}_6\}$ and $\Lambda(v_1) \neq \Lambda(v2)$. Thus, they are not redundant.

\subsubsection{Structure Error}

After merging process, the number of relations between neighbor $\hat{v}_1,\hat{v}_2,...,\hat{v}_J$ and character vertices $v_1,v_2,...,v_I$ have been changed, but the number of relations between character vertices and their neighbor vertices, and the number of neighbors remain unchanged. According to this principle, we use this numbers to define structure error which is the validation criteria of structure merging.

\begin{definition}[\textbf{Structure error}] \label{def:igis se}
In network $G=<V,R,\Phi>$, he character vertices subset is denoted as $\{v_1,v_2,...,v_I\} \subseteq V$, and the neighbor vertices subsets before and after merging are $\hat{v}_1,\hat{v}_2,...,\hat{v}_J \subseteq V$ and $\hat{v}_1,\hat{v}_2,...,\hat{v}_{J'} \subseteq V$, $\forall v_x,v_y \in \{v_1,v_2,...,v_I\}$, then
\begin{equation}\label{equation_1}
  \varepsilon_\xi(v_x,v_y)=\sum_{j=1}^{J}\delta(\hat{v}_j)-\sum_{j'=1}^{J'}\delta'(\hat{v}_{j'})-\frac{\delta(v_X)+\delta(v_y)}{2}
\end{equation}
\end{definition}

In equation~\ref{equation_1}, $\hat{v}_j \in \{\hat{v}_1,\hat{v}_2,...,\hat{v}_J\}$,$\hat{v}_{j'} \in \{\hat{v}_1,\hat{v}_2,...,\hat{v}_{J'}\}$, the numbers of neighbor vertices before and after merging are denoted as $J$ and $J'$. $\delta(v_x)$ and $\delta(v_y)$ severally represent the number of relations between $v_x$ and $v_y$ and their neighbors. $\xi$ represents the type label of social networks and $\Xi$ is the set of type labels, $\xi \in \Xi$. The structure error of $v_x$ and $v_y$ is denoted as $\varepsilon_\xi(v_x,v_y)$.

Based on this notion, we can recognize redundant vertices candidate from social networks according to structure error. If $\varepsilon_\xi(v_x,v_y)=0$, we can regard $v_x$ and $v_y$ as a vertex pair with uniqueness, whereas they are redundant vertices candidates.

\subsection{Algorithm}
We designed redundant vertices candidates screening method in social networks according to the notion of the above. Firstly, we arbitrarily select two character vertices $v_x$ and $v_y$ from networks, and then calculate the number of relations between character vertices and their neighbors. We denote it as \emph{preRelations}. Secondly, based upon merging principle we calculate the number of the relations between them after correct merging, and it is denoted as \emph{postRelations}. Lastly, we calculate structure error of each vertex pairs and put the vertices which have zero value of structure error into redundant vertices candidates set.

\begin{algorithm}
\begin{algorithmic}[1]
\footnotesize
\caption{\bf Candidate redundant vertices screening}
\label{alg:crvs}

\INPUT  social network $G=<V,R,\Phi>$.
\OUTPUT candidate redundant vertices set H.

\STATE Initializing: list $L_1,L_2 \leftarrow V$;

\FOR{$i \leftarrow 1$ to $|V|$}
    \FOR{$j \leftarrow 1$ to $|V|$}
        \IF{the name of $L_1[i] \neq the name of L_2[j]$}
            \IF{$\varepsilon_\xi(L_1[i],L_2[j])=0$}
                \STATE add $L_1[i],L_2[j]$ into H;
            \ENDIF
        \ENDIF
    \ENDFOR
\ENDFOR
\RETURN H;
\end{algorithmic}
\end{algorithm}

\begin{algorithm}
\begin{algorithmic}[1]
\footnotesize
\caption{\bf $\varepsilon_\xi(v_x,v_y)$}
\label{alg:crvs}

\INPUT  two character vertices $v_x$ and $v_y$.
\OUTPUT structure error value $\varepsilon$.

\STATE Initializing: list $L_3 \leftarrow v_x \text{and} v_y$ in $G_\xi$;
\STATE Initializing: list $L_5 \leftarrow$ relations in $G_\xi$ which shared by character vertices $v_x$ and $v_y$;

\FOR{$m \leftarrow 1$ to $|L_3|$}
    \STATE $L_4 \leftarrow$ relations of person node $L_3[m]$
    \FOR{$n \leftarrow 1$ to $|L_4|$}
        \STATE \emph{preRelations} $\leftarrow$ \emph{preRelations} $+1$
    \ENDFOR
\ENDFOR
\FOR{$m \Leftarrow 1$ to $|L_5|$}
    \STATE \emph{postRelations} $\leftarrow$ \emph{postRelations} $+1$
    \STATE $\varepsilon \leftarrow (|L_3|/(|L_3|-1))*$\emph{preRelations}-$($\emph{preRelations}-\emph{postRelations}$)$
\ENDFOR
\RETURN $\varepsilon$;
\end{algorithmic}
\end{algorithm}

\section{Character Uniqueness Measure Based on Activity Path Similarity}
\label{CharacterUniquenessMeasure}

The temporal attributes of semantic relations are composed by start time and end time of activities. We can use them to construct heterogeneous temporal social networks which are consisted by several different types of subnetworks. Each of the subnetworks contains only one type of relations. Vertices similarity is therefor decided by activity relations between character vertices and entity vertices in different subnetworks. As differences of temporal attributes cause differences of relation path, we introduce activity path to describe these network structure. Based on this notion, we quantitatively measure similarity of character vertices by calculating temporal weight of activity paths. After combining all results in each subnetwork, character uniqueness can be measured precisely.

\subsection{5.1	Transaction Activity Network (TAN)}
Let $A=\{\alpha_1,\alpha_2,...,\alpha_n\}$ and $B=\{\beta_1,\beta_2,...,\beta_n\}$ be label sets of vertices types and relation types, $\alpha \in A \bigcap \beta \in B$. Nonempty definite sets $V_\alpha$ and $\hat{V_\alpha}$ denotes character vertices set and entity vertices set respectively. Nonempty relation set is denoted by $R_\beta$. Let $T_\beta$ be temporal attributes set of activities in $G$. The mapping from relations to vertices and temporal attributes is denoted by $\Phi_\beta:R_\beta \leftarrow \{<v_i,\hat{v}_j,\tau_k>|v_i \in V_\alpha \bigcap \hat{v}_j \in \hat{V}_\alpha \bigcap \tau_k \in T_\beta\}$, and its inverse mapping is ${\Phi_\beta}^{-1}$. The mappings of vertices types and relation types are denoted by $\Omega_\alpha:V_\alpha \leftarrow A$ and $\Theta_\beta:R_\beta \leftarrow B$ severally.

\begin{definition}[\textbf{Transaction activity network}] \label{def:igis tan}
A transaction activity network (TAN for short) contains activity information and temporal attributes. It is denoted by $G=<V_\alpha,\hat{V}_\alpha,R_\beta,T_\beta,\Phi_\beta,\Omega_\beta,\Theta_\beta>$, $\alpha \in A \bigcap \beta \in B$.
\end{definition}

\noindent\textbf{Property 1(Heterogeneity)}. In a TAN denoted by $G=<V_\alpha,\hat{V}_\alpha,R_\beta,T_\beta,\Phi_\beta,\Omega_\beta,\Theta_\beta>$, there is $|A| \geq 2 \bigcap |B| \geq 1$.

\noindent\textbf{Property 2(Temporality)}. In a TAN denoted by $G=<V_\alpha,\hat{V}_\alpha,R_\beta,T_\beta,\Phi_\beta,\Omega_\beta,\Theta_\beta>$\\,there is $\tau_k=<\tau^S_k,\tau^E_k> \bigcap \tau_k \in T_\beta$, $\tau_k$ denotes time attribute of $r_k$, $\tau^S_k$ and $\tau^E_k$ denote severally start time and end time of $r_k$.

A large-scale TAN always contains several types of social activities. We can divide it into two or more subnetworks. Each of them contains one type of transaction activities. Let $G_1,G_2,...,G_{|B|}$ be subnetworks with different types of activities. If its relations have temporal attributes and the set is $T_G=\{T_\beta|\beta \in B\}$, We denote $G$ as $G=\{G_\beta|G_\beta=<V_\alpha,\hat{V}_\alpha,R_\beta,T_\beta,\Phi_\beta,\Omega_\beta,\Theta_\beta> \bigcap \alpha \in A \bigcap \beta \in B\}$, and the sets of vertices, relations and types are denoted respectively by $V_G=\{V_\alpha|\alpha \in A\} \bigcup \{\hat{V}_\alpha|\alpha \in A\}$, $R_G=\{R_\beta|\beta \in B\}$, and $\Phi_G=\{\Phi_\beta|\beta \in B\}$. It indicates that $G$ consists of subnetworks $G_1,G_2,...,G_{|B|}$, thus we denote it as $G=\{G_\beta|\beta \in B\}$ simply. Thus, it can be seen that a large-scale TAN contains multi-type vertices and relations, and differences of vertex types lead to differences of relation types~\cite{DBLP:conf/Leicht}. In real world, a TAN always contains several types of social activity, namely there are different types of relations and vertices in a network.

\begin{figure}[thb]
\newskip\subfigtoppskip \subfigtopskip = -0.1cm
\centering
\includegraphics[width=0.9\linewidth]{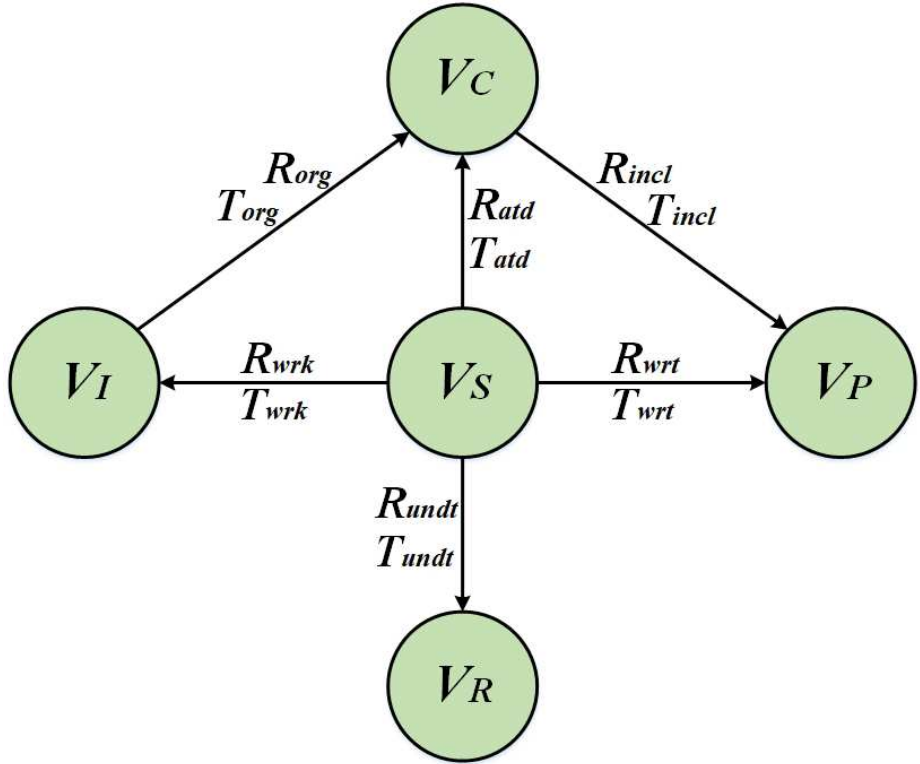}
\vspace{-1mm}
\caption{\small  A heterogeneous TAN }
\label{fig:fig3}
\end{figure}

Figure~\ref{fig:fig3} shows a heterogeneous academic TAN. It contains two types of vertices: scholar vertices $V_S$ and entity vertices $V_I$, $V_C$, $V_P$, $V_R$ which represent \emph{institution}, \emph{conference}, \emph{publication} and \emph{research project}. Due to differences of academic activities, there are different relations between vertices, such as write relation between scholars and papers, participation relation between scholars and conferences, etc. We use $R_\text{org}$, $R_\text{atd}$, $R_\text{incl}$, $R_\text{wrk}$, $R_\text{wrt}$ and $R_\text{undt}$ to denote six types of relations (organize, attend, included, work at, write and undertake) and $T_\text{org}$, $T_\text{atd}$, $T_\text{incl}$, $T_\text{wrk}$, $T_\text{wrt}$ and $T_\text{undt}$ denote temporal attribute set.

\subsection{Transaction Activity Path (TAP)}
In a TAN, transaction activity paths (TAPs for short) are relative to topology of it. We regard character vertices and entity vertices respectively as master vertices and their neighbor vertices, and then we can describe TAPs. A TAP is a path which goes through a pair of character vertices and one entity vertices and the relations between them. From one master vertex to another, there is one or more TAP through their common neighbors, and they contain semantics and temporal attributes of original transaction records.

Let character and neighbor vertices be $v_x, v_y \in V_\alpha$ and $\hat{v}_z \in \hat{V}_\alpha$ respectively, $X_1=\{1,2,3,...,\chi_1,...,|X_1|\}$ and $X_2=\{1,2,3,...,\chi_2,...,|X_2|\}$ are two label sets of relations in relation sets $R_{xz}$ and $R_{yz}$ and

\begin{equation}\label{equation_2}
R_{xz}=\{r_{\chi_1}|\chi_1 \in X_1\}={\Phi_\beta}^{-1}\{<v_x,\hat{v_z},\tau_{\chi_1}>|v_x \in V_\alpha \bigcap \hat{v}_z \in \hat{V}_\alpha \bigcap \tau_{\chi_1} \in T_\beta\}
\end{equation}

\begin{equation}\label{equation_3}
R_{yz}=\{r_{\chi_2}|\chi_2 \in X_2\}={\Phi_\beta}^{-1}\{<v_y,\hat{v_z},\tau_{\chi_2}>|v_y \in V_\alpha \bigcap \hat{v}_z \in \hat{V}_\alpha \bigcap \tau_{\chi_2} \in T_\beta\}
\end{equation}

\begin{definition}[\textbf{Transaction activity path}] \label{def:igis tan}
In a TAN $G=<V_\alpha,\hat{V}_\alpha,R_\beta,T_\beta,\Phi_\beta,\Omega_\beta,$\\$\Theta_\beta>$, let $v_x$ or $v_y$ be start vertex, a path which begin at $v_x$, and go through neighbor vertex $\hat{v}_z$ and then end at $v_y$ is called as transaction activity path. It denoted by $p(v_x\hat{v}_zv_y)_{\chi_1\chi_2}$. The set of TAPs between $v_x$ and $v_y$ is denoted by $P_{xy}=\{p(v_x\hat{v}_zv_y)_{\chi_1\chi_2}|z \in Z \bigcap \chi_1 \in X_1 \bigcap \chi_2 \in X_2\}$.
\end{definition}

\noindent\textbf{Property 1}. Let $|P_{xy}|$ be the number of TAPs in set $P_{xy}$, $|P_{xy}|=|X_1|*|X_2|$.


\textbf{Instance 1}. Figure 4 shows that in a TAN $G_\beta=<V_\text{person},\hat{V}_\text{club},R_\beta,T_\beta,\Phi_\beta,\Omega_\beta,\Theta_\beta$\\$>$, the sets of character vertices and their neighbor vertices are $V_\beta=\{v_1,v_2\}$ and $\hat{V}_\beta=\{\hat{v}_1,\hat{v}_2\}$, $R_\beta=\{r_1,r_2,r_3,r_4,r_5\}$ and $T_\beta=\{\tau_1,\tau_2,\tau_3,\tau_4,\tau_5\}$ are the sets of relations and temporal attributes. We can find that $R_{11}=\{r_1,r_2\}$,$R_{12}=\{r_3\}$,$R_{21}=\{r_4\}$,$R_{22}=\{r_5\}$. Evidently though, there are two activity paths from vertex $v_1$ to $v_2$ through neighbor vertex $\hat{v}_1$, and we denote them by $p(v_1\hat{v}_1v_2)_{14}$ and $p(v_1\hat{v}_1v_2)_{24}$. Similarly, we denote the path through neighbor $\hat{v}_2$ by $p(v_1\hat{v}_2v_2)_{35}$.

\subsection{Character Uniqueness Measure}
Owing to temporal attributes of relations, we can define and calculate temporal weight of relations and TAPs, which reflect temporal characteristics of transaction activity networks. Based on temporal weight we can calculate TAP similarity to measure similarity degree of character vertices pairs. The similarity threshold is a filter to screen out unique vertices so that we can get redundant vertices set.

\subsubsection{Temporal weight calculation}
In a transaction activity network, temporal weights of relations are decided by start time and end time, while temporal weights of TAPs are decided by the former. Based on time attribute $\tau_k=<\tau^S_k,\tau^E_k>$, we can use the following equation to calculate temporal weight of $r_k$:

\begin{equation}\label{equation_4}
W^z_k=(Now+1-\tau^S_k)*(\tau^E_k+1-\tau^S_k)
\end{equation}

Now denotes current data, k denotes label of relations and $z$ is label of neighbor vertex $\hat{v}_z$. The following equation is the temporal weight of TAPs:

\begin{equation}\label{equation_4}
W(p(v_x\hat{v}_zv_y)_{\chi_1\chi_2})=W^z_{\chi_1}*W^z_{\chi_2}
\end{equation}

The temporal weight of relations reflects the start time and end time, as well as the duration of relations. Apparently, the temporal weight of TAPs contains all of this information since TAPs are consisted by two relations. The weight is decided by the temporal attributes of relations.

\subsubsection{Transaction activity path similarity}
In a transaction activity network $G_\beta$, let character vertices and entity vertex be $v_x, v_y \in V_\alpha$ and $\hat{v}_z \in \hat{V}_\alpha$ which is the neighbor of $v_x$ and $v_y$. The TAP sets are denoted by $P_{xy}=\{p(v_x\hat{v}_zv_y)_{\chi_1\chi_2}|z \in Z \bigcap \chi_1 \in X_1 \bigcap \chi_2 \in X_2\}$, $P_{xx}=\{p(v_x\hat{v}_zv_x)_{\chi_1\chi_1}|z \in Z \bigcap \chi_1 \in X_1 \bigcap \chi_1 \in X_1\}$ and $P_{yy}=\{p(v_y\hat{v}_zv_y)_{\chi_2\chi_2}|z \in Z \bigcap \chi_2 \in X_2 \bigcap \chi_2 \in X_2\}$. $P_{xy}$, $P_{xx}$ and $P_{yy}$ represent three types of paths respectively. They have three different structures:

\begin{figure}[thb]
\newskip\subfigtoppskip \subfigtopskip = -0.1cm
\centering
\includegraphics[width=0.6\linewidth]{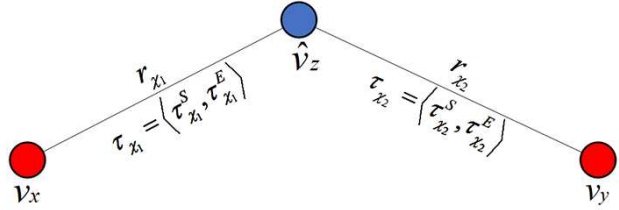}
\vspace{-1mm}
\caption{\small The first type of TAP }
\label{fig:fig5}
\end{figure}

Figure~\ref{fig:fig5} shows the first type of TAP between $v_x$ and $v_y$. These paths begin from $v_x$ then go through relations $r_{\chi_1}$, neighbor $\hat{v}_z$ and relation $r_{\chi_2}$, and end to $v_y$. In a network, all of the TAPs between different two vertices are this type. The second and third types are showed in Figures. Both of them begin from one vertex ($v_x$ or $v_y$) and end to the same vertex, and they are through the same relation twice.

\begin{definition}[\textbf{SimTAP}] \label{def:igis simtap}
SimTAP is the similarity between two vertices $v_x$ and $v_y$. It is decided by structure and temporal weight of TAPs between $v_x$ and $v_y$. The definition formula of SimTAP is following:
\begin{equation}\label{equation_6}
SimTAP_\beta(v_x,v_y)=\frac{2*W(P_{xy})}{W(P_{xx})+W(P_{yy})}
\end{equation}
\end{definition}

In this formula, $W(P_{xy})$, $W(P_{xx})$ and $W(P_{yy})$ denote the temporal weight sums of these three types of TAPs. We use the following formulas to calculate these weights:

\begin{equation}\label{equation_7}
W(P_{xy})=\sum_{z=1}^{|Z|}\sum_{\chi_1=1}^{|X_1|}\sum_{\chi_2=1}^{|X_2|}X^z_{\chi_1}*W^z_{\chi_2}
\end{equation}

\begin{equation}\label{equation_8}
W(P_{xx})=\sum_{z=1}^{|Z|}\sum_{\chi_1=1}^{|X_1|}\sum_{\chi_1=1}^{|X_1|}X^z_{\chi_1}*W^z_{\chi_1}
\end{equation}

\begin{equation}\label{equation_9}
W(P_{yy})=\sum_{z=1}^{|Z|}\sum_{\chi_2=1}^{|X_2|}\sum_{\chi_2=1}^{|X_2|}X^z_{\chi_2}*W^z_{\chi_2}
\end{equation}

$W^z_{\chi_1}$ and $W^z_{\chi_2}$ are weights of relations between $v_x$ and $v_y$.

Generally, a transaction activity network contains several subnetworks. In order to measure the similarity of all characters, we need to add all similarity values in each subnetwork and then calculate arithmetic mean. Let a TAN be $G=\{G_\beta|\beta \in B\}$, we calculate similarity of vertices pair $v_x$ and $v_y$ in $G_\beta$, then we get the TAPs similarity set $\{SimTAP_\beta(v_x,v_y)|\beta \in B\}$. After that we calculate the arithmetic mean in $G$. The formula is following:

\begin{equation}\label{equation_10}
SimTAP(v_x,v_y)=\frac{1}{|B|}\sum_{\beta=1}^{|B|}SimTAP_\beta(v_x,v_y)
\end{equation}

In the formula, $|B|$ is the number of subnetworks in $G$, $SimTAP(v_x,v_y)$ is the TAP similarity of $v_x$ and $v_y$.

\subsubsection{Character uniqueness measure}
$SimTAP(v_x,v_y)$ can measure uniqueness of characters quantitatively in TANs. The value of it is larger, the similarity between character vertices is greater, and vice versa. According to this idea, we proposed uniqueness measurement of characters: after $SimTAP(v_x,v_y)$ calculating, we set character uniqueness threshold $\theta$ based on features of networks and data-analytic requirements to screen out the results. If $SimTAP(v_x,v_y) < \theta$, we regard $v_x$ and $v_y$ as unique characters. While if $SimTAP(v_x,v_y) \geq \theta$, vertices $v_x$ and $v_y$ have high similarity, which indicates that we need to merge these vertices and their shared relations.

\textbf{Instance 2}. In a transaction activity network $G=\{G_S,G_W,G_R,G_C\}$, the type sets of vertices and relations are respectively denoted by $A$ and $B$=\{Study,Work,Re-\\search,Coauthor\}. There are 10 character vertices in this network and the values of similarity of them are showed in table~\ref{table_1}.

\begin{table}
\centering
\caption{The results of transaction activity similarity}
\arrayrulecolor{black}
\begin{tabular}{!{\color{black}\vrule}l!{\color{black}\vrule}l!{\color{black}\vrule}l!{\color{black}\vrule}l!{\color{black}\vrule}l!{\color{black}\vrule}l!{\color{black}\vrule}l!{\color{black}\vrule}}
\hline
 \textit{$v_x$}  &  \textit{$v_y$}  &  \textit{$G_S$}       &  \textit{$G_W$}       &  \textit{$G_R$}       &  \textit{$G_C$}       &  \textit{$G$}   \\
\hline
Long Chen     & Jay Liu       & 0.0000 & 0.4235 & 1.0000 & 1.0000 & 0.6059        \\
\hline
Hao Feng      & Wei Zhang     & 0.8661 & 1.0000 & 1.0000 & 0.0000 & 0.7165        \\
\hline
Jing Xu       & Jiang Zhou    & 1.0000 & 1.0000 & 1.0000 & 0.0000 & 0.7500        \\
\hline
Faye Wu       & Fei X. Wu     & 1.0000 & 1.0000 & 1.0000 & 1.0000 & 1.0000        \\
\hline
Faye Wu       & Ron Xiao      & 0.5219 & 0.0000 & 1.0000 & 1.0000 & 0.6305        \\
\hline
Faye Wu       & Pan Zhang     & 0.5219 & 0.0000 & 1.0000 & 1.0000 & 0.6305        \\
\hline
Fei X. Wu     & Ron Xiao      & 0.5219 & 0.0000 & 1.0000 & 1.0000 & 0.6305        \\
\hline
Fei X. Wu     & Pan Zhang     & 0.5219 & 0.0000 & 1.0000 & 1.0000 & 0.6305        \\
\hline
Ron Xiao      & Pan Zhang     & 1.0000 & 0.0000 & 1.0000 & 1.0000 & 0.7500        \\
\hline
\end{tabular}
\arrayrulecolor{black}
\label{table_1}
\end{table}

In This Table, columns $v_x$ and $v_y$ are name of characters, $G_S$, $G_W$, $G_R$, $G_C$  indicate the similarity of vertices pairs in these four subnetworks, and $G$ denote the similarity in $G$. After setting the threshold $\theta=0.70$, we can find that there are four similarity values larger than $\theta$: $SimTAP(v_{Hao Feng}, v_{Wei Zhang})$, $SimTAP(v_{Jing Xu}, v_{Jiang Zhou})$, $SimTAP(v_{Faye Wu}, v_{Fei X.Wu})$ and $SimTAP(v_{Ron Xiao}, v_{Pan Zhang})$. It indicates that these characters are remarkably similar, and they do not have uniqueness. Besides, the similarity of character \emph{Long Chen} and \emph{Jay Liu} is smaller than $\theta$, namely $SimTAP$\\$(v_{Long Chen}, v_{Jay Liu})<\theta$, thus they have uniqueness. This shows that we can screen out the character vertices which have uniqueness by calculating similarity $SimTAP$\\$(v_x, v_y)$ and setting threshold $\theta$.

\subsection{Algorithm Design}
We designed TAPs similarity algorithm based on above-mentioned theories. At first, we get the relation lists of vertices pair $v_x$ and $v_y$ from each subnetwork $G_\beta$, and then calculate the temporal weight of transaction activity paths. Second, we calculate the transaction activity network similarity $SimTAP\beta$ of $v_x$ and $v_y$, then calculate arithmetic mean of similarity $SimTAP$ in network $G$. After traversing all vertices pairs in candidate redundant vertices set H and get their similarity, we set threshold $\theta$ and compare it with each similarity. We regard the vertices whose similarity is larger than $\theta$ as redundant vertices and put them into redundant vertices set $\widetilde{H}$. The vertices whose similarity is smaller than $\theta$ are regarded as unique vertices and they must remain in network.

\begin{algorithm}
\begin{algorithmic}[1]
\footnotesize
\caption{\bf Redundant vertices screening in heterogeneous network $G$}
\label{alg:rvshnG}

\INPUT  candidate redundant vertices set H.
\OUTPUT redundant vertices set H.

\STATE Initializing: list $L_1,L_2$ by candidate redundant vertices set H;

\FOR{$i \leftarrow 1$ to $|L_1|$}
    \FOR{$j \leftarrow 1$ to $|L_2|$}
        \IF{the name of $L_1[i] = the name of L_2[j]$}
            \IF{$SimTAP(L_1[i],L_2[j]) \geq \theta$}
                \IF{$L_1[i] \in H_i$ or $L_2[j] \in H_j$}
                    \STATE Insert $L_1[i]$ or $L_2[j]$ into $\text{H}_i$;
                \ELSE
                    \STATE Insert $L_1[i]$, $L_2[j]$ into $\text{H}_{i+1}$;
                \ENDIF
            \ENDIF
        \ENDIF
    \ENDFOR
\ENDFOR
\RETURN H;
\end{algorithmic}
\end{algorithm}

\begin{algorithm}
\begin{algorithmic}[1]
\footnotesize
\caption{\bf $SimTAP_\beta(v_x,v_y)$}
\label{alg:rvshnG}

\INPUT  Two character vertices $v_x$ and $v_y$.
\OUTPUT value of path similarity of $v_x$ and $v_y$ in $G_\beta$.

\STATE Initializing: list $R_1$ by relation list of $v_x$ in $G_\beta$;
\STATE Initializing: list $R_2$ by relation list of $v_y$ in $G_\beta$;

\FOR{$k_1 \leftarrow 1$ to $|R_1|$}
    \FOR{$k_2 \leftarrow 1$ to $|R_2|$}
        \IF{$v_z$ of $R_1[k_1]=v_z$ of $R_2[k_2]$}
            \STATE $W(p(v_x\hat{v}_zv_y)_{\chi_1\chi_2}) \leftarrow W^z_{\chi_1} * W^z_{\chi_2}$
        \ENDIF
    \ENDFOR
\ENDFOR
\FOR{$k_1 \leftarrow 1$ to $|R_1|$}
    \FOR{$k_2 \leftarrow 1$ to $|R_1|$}
        \IF{$v_z$ of $R_1[k_1]=v_z$ of $R_1[k_2]$}
            \STATE $W(p(v_x\hat{v}_zv_x)_{\chi_1\chi_1}) \leftarrow W^z_{\chi_1} * W^z_{\chi_1}$
        \ENDIF
    \ENDFOR
\ENDFOR
\FOR{$k_1 \leftarrow 1$ to $|R_2|$}
    \FOR{$k_2 \leftarrow 1$ to $|R_2|$}
        \IF{$v_z$ of $R_2[k_1]=v_z$ of $R_2[k_2]$}
            \STATE $W(p(v_y\hat{v}_zv_y)_{\chi_2\chi_2}) \leftarrow W^z_{\chi_2} * W^z_{\chi_2}$
        \ENDIF
    \ENDFOR
\ENDFOR
\STATE $SimTAP_\beta \leftarrow (2*W(p(v_x\hat{v}_zv_y))/(W(p(v_x\hat{v}_zv_x)+W(p(v_y\hat{v}_zv_y))$
\RETURN $SimTAP_\beta$;
\end{algorithmic}
\end{algorithm}

\begin{algorithm}
\begin{algorithmic}[1]
\footnotesize
\caption{\bf $SimTAP(v_x,v_y)$}
\label{alg:rvshnG}

\INPUT  two character vertices $v_x$ and $v_y$.
\OUTPUT value of path similarity of $v_x$ and $v_y$ in $G$.

\STATE Initializing: $SimTAP \leftarrow 0$;

\FOR{$\beta \leftarrow 1$ to $|B|$}
    \STATE $SimTAP \leftarrow SimTAP + SimTAP_\beta(v_x,v_y)$
\ENDFOR
\STATE $SimTAP \leftarrow SimTAP / |B|$
\RETURN $SimTAP(v_x,v_y)$;
\end{algorithmic}
\end{algorithm}

\subsection{Experiment and analysis}
The multimedia dataset for academic transaction networks building contains texts, images and videos concerning proposals, papers, award certificates and videos of academic conference. In the experiment of this paper, we extract academic activity transaction data from 724 proposals of Natural Science Foundation of China (\emph{NSFC})~\cite{DBLP:conf/ShanmeiTang}, which are texts in Chinese only, and then established a transaction database. After that, we import these data into graph database \emph{Neo4J} and then construct transaction activity networks which contained 598 vertices. We mine academic relationship between scholars and then build academic networks. On this basis, we calculate structure error of character vertices and then give the visual presentation of network~\cite{DBLP:conf/WenjieXiao}. Based on the results of structure error calculation, we get vertices from redundant vertices set H and calculated SimTAP of each vertices pair.

\subsubsection{Evaluation for structure error}
Our academic transaction activity networks contain 589 scholars’ academic transaction information. The first step was extracting academic activity data from transaction database, and then imported them into graph database. We construct four types of activity networks, they were education experience network, work experience network, project cooperation network and co-author network. After that, we build academic network G based on them. Figure 3 shows this network.

In this network, we calculate structure error of each vertices pair and screen out the vertices with 0 structure error. Table~\ref{table_2} shows partial results.

\begin{table}
\centering
\caption{Structure error of vertices in $G$}
\arrayrulecolor{black}
\begin{tabular}{!{\color{black}\vrule}l!{\color{black}\vrule}l!{\color{black}\vrule}l!{\color{black}\vrule}}
\hline
 \textit{$v_x$}  &  \textit{$v_y$}  &  \textit{$\varepsilon_G$}(\textit{$v_x$},\textit{$v_y$})   \\
\hline
Faye Wu       & Fei Wu        & 0.000                                   \\
\hline
ShaoJia Zhu   & ShaoNan Zhu   & 0.000                                   \\
\hline
Ruifeng Fan   & Chi Zhang     & 0.250                                   \\
\hline
Xiaojie Liu   & Chao Zhang    & 0.500                                   \\
\hline
Kang Du       & Lin Guo       & 0.500                                   \\
\hline
Lijuan Liu    & Guanghua Zhao & 0.625                                   \\
\hline
Yafei Hou     & Yue Wang      & 0.750                                   \\
\hline
Shengmin Fu   & Yanni Peng    & 0.750                                   \\
\hline
Zeo Zhou      & Ze Zhong      & 0.750                                   \\
\hline
Bin Du        & Binbin Gao    & 0.875                                   \\
\hline
\end{tabular}
\arrayrulecolor{black}
\label{table_2}
\end{table}

\begin{table}
\centering
\caption{Education experience info of Faye Wu AND Fei Wu}
\arrayrulecolor{black}
\begin{tabular}{!{\color{black}\vrule}l!{\color{black}\vrule}l!{\color{black}\vrule}l!{\color{black}\vrule}l!{\color{black}\vrule}}
\hline
Name    & Institute              & Start Time & End Time  \\
\hline
Faye Wu & Hubei Minzu Univ.      & 1992       & 1996      \\
\hline
Faye Wu & Guangzhou Normal Univ. & 1997       & 1999      \\
\hline
Faye Wu & Jinan Univ.            & 2000       & 2005      \\
\hline
Fei Wu  & Hubei Minzu Univ.      & 1992       & 1996      \\
\hline
Fei Wu  & Guangzhou Normal Univ. & 1997       & 1999      \\
\hline
Fei Wu  & Jinan Univ.            & 2000       & 2000      \\
\hline
\end{tabular}
\arrayrulecolor{black}
\label{table_3}
\end{table}

\begin{table}
\centering
\caption{Work experience info of Faye Wu AND Fei Wu}
\arrayrulecolor{black}
\begin{tabular}{!{\color{black}\vrule}l!{\color{black}\vrule}l!{\color{black}\vrule}l!{\color{black}\vrule}l!{\color{black}\vrule}}
\hline
Name    & Employer            & Start Time & End Time  \\
\hline
Faye Wu & Central South Univ. & 2010       & 2012      \\
\hline
Fei Wu  & Central South Univ. & 2010       & 2012      \\
\hline
\end{tabular}
\arrayrulecolor{black}
\label{table_4}
\end{table}

In Table~\ref{table_2}, fields $v_x$ and $v_y$ denote two vertices, field denotes value of structure error of these vertices in $G$. We can find that structure error of vertices pairs \emph{Faye Wu} and \emph{Fei Wu}, \emph{Shaojia Zhu} and \emph{Shaonan Zhu} are zero. Therefore, these two vertices are regarded as redundant vertices candidates. We can find their structure features in Figure. Four highlighted character vertices are \emph{Faye Wu}, \emph{Fei Wu}, \emph{Shaojia Zhu} and \emph{Shaonan Zhu}. These two highlighted subnetworks illustrate that the two vertex pairs have same neighbors respectively.

In order to analyze our method deeply, we extract academic activity information from the database. Table~\ref{table_3} to Table~\ref{table_6} show academic activity information of \emph{Faye Wu} and \emph{Fei Wu}.

\begin{table}
\centering
\caption{Project information of Faye Wu AND Fei Wu}
\arrayrulecolor{black}
\begin{tabular}{!{\color{black}\vrule}l!{\color{black}\vrule}l!{\color{black}\vrule}l!{\color{black}\vrule}l!{\color{black}\vrule}}
\hline
Name    & Project                                                                                                        & Start Time & End Time  \\
\hline
Faye Wu & \begin{tabular}[c]{@{}l@{}} Control Policy Research in Parabolic~\\Distributed Parameter Systems \end{tabular} & 2010       & 2012      \\
\hline
Fei Wu  & \begin{tabular}[c]{@{}l@{}} Control Policy Research in Parabolic~\\Distributed Parameter Systems \end{tabular} & 2010       & 2012      \\
\hline
\end{tabular}

\arrayrulecolor{black}
\label{table_5}
\end{table}

\begin{table}
\centering
\caption{Publication info of Faye Wu AND Fei Wu}
\arrayrulecolor{black}
\begin{tabular}{!{\color{black}\vrule}l!{\color{black}\vrule}l!{\color{black}\vrule}l!{\color{black}\vrule}l!{\color{black}\vrule}}
\hline
Name    & Publication                                                                                                                                   & Start Time & End Time  \\
\hline
Faye Wu & \begin{tabular}[c]{@{}l@{}} Adaptive Control Synchronization~\\Approach Research of Unified Chaotic Systems \end{tabular}                     & 2000       & 2000      \\
\hline
Faye Wu & \begin{tabular}[c]{@{}l@{}} Linear and Nonlinear Feedback Synchronization\\~and Performance Research in Discrete Chaotic System \end{tabular} & 2004       & 2004      \\
\hline
Fei Wu  & \begin{tabular}[c]{@{}l@{}} Adaptive Control Synchronization Approach Research~\\of Unified Chaotic Systems \end{tabular}                     & 2000       & 2000      \\
\hline
Fei Wu  & \begin{tabular}[c]{@{}l@{}} Linear and Nonlinear Feedback Synchronization and\\~Performance Research in Discrete Chaotic System \end{tabular} & 2004       & 2004      \\
\hline
\end{tabular}

\arrayrulecolor{black}
\label{table_6}
\end{table}

We can find that \emph{Faye Wu} and \emph{Fei Wu} studied in the same school over the same period. Likewise, they have the same experience on the aspects of work, project and publication. Namely their experience of academy is selfsame. Thus, \emph{Faye Wu} or \emph{Fei Wu} is not unique, which is redundant information.

In Figure, vertices \emph{Shaojia Zhu} and \emph{Shaonan Zhu} own the same neighbors. Similarly, we extract their activity information.

From Table~\ref{table_7} to Table~\ref{table_10} we can see that vertices \emph{Shaojia Zhu} and \emph{Shaonan Zhu} studied in the same universities and employed by the same employer but the periods are different. That means their education and work experience is different. The difference between \emph{Shaojia Zhu} and \emph{Shaonan Zhu} is caused by the difference of temporal attributes. Therefore, both of them are unique and they do not contain redundant information.

\begin{table}
\centering
\caption{Education experience info of Shaojia Zhu AND Shaonan Zhu}
\arrayrulecolor{black}
\begin{tabular}{!{\color{black}\vrule}l!{\color{black}\vrule}l!{\color{black}\vrule}l!{\color{black}\vrule}l!{\color{black}\vrule}}
\hline
Name        & Institute                     & Start Time & End Time  \\
\hline
ShaoJia Zhu & Changsh Univ. of Sci and Tech & 1990       & 1994      \\
\hline
ShaoJia Zhu & Central South Univ.           & 1996       & 1999      \\
\hline
ShaoJia Zhu & Zejiang Univ.                 & 2000       & 2004      \\
\hline
ShaoNan Zhu & Changsh Univ. of Sci and Tech & 2000       & 2004      \\
\hline
ShaoNan Zhu & Central South Univ.           & 2004       & 2007      \\
\hline
ShaoNan Zhu & Zejiang Univ.                 & 2008       & 2012      \\
\hline
\end{tabular}
\arrayrulecolor{black}
\label{table_7}
\end{table}

\begin{table}
\centering
\caption{Work experience info of Shaojia Zhu AND Shaonan Zhu}
\arrayrulecolor{black}
\begin{tabular}{!{\color{black}\vrule}l!{\color{black}\vrule}l!{\color{black}\vrule}l!{\color{black}\vrule}l!{\color{black}\vrule}}
\hline
Name        & Employer                    & Start Time & End Time  \\
\hline
ShaoJia Zhu & Hunan Univ. of Sci and Tech & 2004       & 2014      \\
\hline
ShaoNan Zhu & Hunan Univ. of Sci and Tech & 2012       & 2014      \\
\hline
\end{tabular}
\arrayrulecolor{black}
\label{table_8}
\end{table}

\begin{table}
\centering
\caption{Project information of Shaojia Zhu AND Shaonan Zhu}
\arrayrulecolor{black}
\begin{tabular}{!{\color{black}\vrule}l!{\color{black}\vrule}l!{\color{black}\vrule}l!{\color{black}\vrule}l!{\color{black}\vrule}}
\hline
Name        & Project                                                                                         & Start Time & End Time  \\
\hline
ShaoJia Zhu & \begin{tabular}[c]{@{}l@{}} Decimal Encryption Technology Research\\~Based on AES \end{tabular} & 2012       & 2013      \\
\hline
ShaoNan Zhu & \begin{tabular}[c]{@{}l@{}} Decimal Encryption Technology Research~\\Based on AES \end{tabular} & 2012       & 2013      \\
\hline
\end{tabular}
\arrayrulecolor{black}
\label{table_9}
\end{table}

\begin{table}
\centering
\caption{Publication info of Shaojia Zhu AND Shaonan Zhu}
\arrayrulecolor{black}
\begin{tabular}{!{\color{black}\vrule}l!{\color{black}\vrule}l!{\color{black}\vrule}l!{\color{black}\vrule}l!{\color{black}\vrule}}
\hline
Name        & Publication                                                                                                             & Start Time & End Time  \\
\hline
ShaoJia Zhu & \begin{tabular}[c]{@{}l@{}} Node Behavior Prediction and Optimized~\\Routing Algorithm in Mobile Networks \end{tabular} & 2013       & 2013      \\
\hline
ShaoNan Zhu & \begin{tabular}[c]{@{}l@{}} Node Behavior Prediction and Optimized~\\Routing Algorithm in Mobile Networks \end{tabular} & 2013       & 2013      \\
\hline
\end{tabular}

\arrayrulecolor{black}
\label{table_10}
\end{table}

The results indicate that character vertices which have same neighbors may do not contain exact same social activity information. These vertices are redundant candidates and among them there are some vertices with uniqueness. But we cannot recognize them by structure error. On the contrary, we can only screen out vertices whose structure error is not zero. They exactly have uniqueness. Above all, we need recognize character uniqueness ulteriorly.

\subsubsection{TAPs similarity calculation}

\begin{table}
\centering
\caption{The results of transaction activity similarity}
\arrayrulecolor{black}
\begin{tabular}{!{\color{black}\vrule}l!{\color{black}\vrule}l!{\color{black}\vrule}l!{\color{black}\vrule}l!{\color{black}\vrule}l!{\color{black}\vrule}l!{\color{black}\vrule}l!{\color{black}\vrule}}
\hline
 \textit{$v_x$}  &  \textit{$v_y$}  &    \textit{$G_S$}    &    \textit{$G_W$}    &    \textit{$G_R$}    &    \textit{$G_C$}    &  \textit{$G$}   \\
\hline
Faye Wu       & Fei Wu        & 0.0000 & 0.4235 & 1.0000 & 1.0000 & 0.6059        \\
\hline
ShaoJia Zhu   & ShaoNan Zhu   & 0.8661 & 1.0000 & 1.0000 & 0.0000 & 0.7165        \\
\hline
Jie Gao       & Di Feng       & 1.0000 & 1.0000 & 1.0000 & 0.0000 & 0.7500        \\
\hline
XinHua Zou    & XingXing Zou  & 1.0000 & 1.0000 & 1.0000 & 1.0000 & 1.0000        \\
\hline
Zhe Feng      & Kang Du       & 0.5219 & 0.0000 & 1.0000 & 1.0000 & 0.6305        \\
\hline
\end{tabular}
\arrayrulecolor{black}
\label{table_11}
\end{table}

We first calculate the similarity of vertices pair in an academic network containing 589 characters. After setting θ as 0.70, we screen out the vertices whose SimTAP is higher than $\theta$. The results are showed in Table~\ref{table_11}.

In this table, we find that the value of similarity of vertices pair \emph{Faye Wu} and \emph{Fei Wu} is 1.0000, which indicates that their academic activity information is identical. That means the similarity between them has been maximized. The similarities of \emph{Jia Gao} and \emph{Di Feng}, \emph{Xinhua Zou} and \emph{Xingxing Zou}, and \emph{Zhe Feng} and \emph{Kang Du} are 0.7450, 0.7463,0.8546 and 0.7500.

\subsubsection{Regression analysis}
We chose the vertices whose similarity in subnetworks is zero and extracted their transaction information from database. It is showed in Table~\ref{table_12},Table~\ref{table_13},Table~\ref{table_14}.

\begin{table}
\centering
\caption{Education info of Jia Gao and Di Feng}
\arrayrulecolor{black}
\begin{tabular}{!{\color{black}\vrule}l!{\color{black}\vrule}l!{\color{black}\vrule}l!{\color{black}\vrule}l!{\color{black}\vrule}l!{\color{black}\vrule}}
\hline
Name    & Institute                      & Start Time & End Time & Degree  \\
\hline
Jie Gao & Qingdao University             & 2005       & 2008     & Ph.D.   \\
\hline
Jie Gao & Heilongjiang University        & 2002       & 2005     & M.Sc.   \\
\hline
Jie Gao & Zhengzhou University           & 1998       & 2002     & B.Sc.   \\
\hline
Di Feng & Peking University              & 1999       & 2004     & Ph.D.   \\
\hline
Di Feng & Nanjing Agriculture University & 1996       & 1999     & M.Sc.   \\
\hline
Di Feng & Guangxi University             & 1992       & 1996     & B.Sc.   \\
\hline
\end{tabular}
\arrayrulecolor{black}
\label{table_12}
\end{table}

\begin{table}
\centering
\caption{Education info of Xinhua Zou and Xingxing Zou}
\arrayrulecolor{black}
\begin{tabular}{!{\color{black}\vrule}l!{\color{black}\vrule}l!{\color{black}\vrule}l!{\color{black}\vrule}l!{\color{black}\vrule}l!{\color{black}\vrule}}
\hline
Name         & Institute                    & Start Time & End Time & Degree  \\
\hline
XinHua Zou   & Central South University     & 2002       & 2006     & Ph.D.   \\
\hline
XinHua Zou   & Hunan University of Medicine & 1999       & 2001     & M.Sc.   \\
\hline
XinHua Zou   & Chang’an University          & 1995       & 1999     & B.Sc.   \\
\hline
XingXing Zou & Peking University            & 2005       & 2008     & Ph.D.   \\
\hline
XingXing Zou & Hunan University             & 2002       & 2005     & M.Sc.   \\
\hline
XingXing Zou & Ocean University of China    & 1998       & 2002     & B.Sc.   \\
\hline
\end{tabular}
\arrayrulecolor{black}
\label{table_13}
\end{table}

\begin{table}
\centering
\caption{Publication info of Zhe Feng and Kang Du}
\arrayrulecolor{black}
\begin{tabular}{!{\color{black}\vrule}l!{\color{black}\vrule}l!{\color{black}\vrule}l!{\color{black}\vrule}l!{\color{black}\vrule}}
\hline
Name     & Institute                                                                                             & Start Time & End Time  \\
\hline
Zhe Feng & \begin{tabular}[c]{@{}l@{}} Lyapunov Exponent Algorithm Design and~\\Implementation \end{tabular}     & 1992       & 1996      \\
\hline
Kang Du  & \begin{tabular}[c]{@{}l@{}} Cell Image Separation Algorithm Based on~\\Contour-stripped \end{tabular} & 1997       & 1999      \\
\hline
\end{tabular}
\arrayrulecolor{black}
\label{table_14}
\end{table}

We can see from the Table~\ref{table_12} that \emph{Jia Gao} and \emph{Di Feng} studied in three different universities. Likewise, in Table~\ref{table_13}, \emph{Xinghua Zou} and \emph{Xingxing Zou} studied in different colleges as well. In Table~\ref{table_14}, the publications of \emph{Zhe Feng} and \emph{Kang Du} are entirely different. These situations indicate that these three pairs of character are different in education and publication activities. It leads to differences of academic relationship between them. However, high similarity of other types of academic activities leads to high value of SimTAP of these characters. It is even higher than threshold $\theta$ so that these characters cannot be screen out from networks. This situation adverse impact character uniqueness identification. This problem can be solved by calculating TAPs similarity and screening out redundant character vertices from social networks.

Based on experiment results in Section 4, we calculated similarity of candidate redundant vertices in H. The results of structure error calculation are showed in Table~\ref{table_15}.

\begin{table}
\centering
\caption{The results of structure error calculation}
\arrayrulecolor{black}
\begin{tabular}{!{\color{black}\vrule}l!{\color{black}\vrule}l!{\color{black}\vrule}l!{\color{black}\vrule}}
\hline
 \textit{$v_x$}  &  \textit{$v_y$}  &  \textit{$\varepsilon G$($v_x$, $v_y$)}   \\
\hline
Faye Wu       & Fei Wu        & 0.0000                 \\
\hline
Shaojia Zhu   & Shaonan Zhu   & 0.0000                 \\
\hline
\end{tabular}
\arrayrulecolor{black}
\label{table_15}
\end{table}

We got the redundant vertices set $H=\{v_\text{Faye Wu},v_\text{Fei Wu},v_\text{Shaojia Zhu},v_\text{Shaonan Zhu}\}$ after structure error calculation, and then calculated the similarity of these four vertices. $\theta$ was set as 0.80, the results are showed in Table 16.

\begin{table}
\centering
\caption{The vertices are screened out by $\theta$}
\arrayrulecolor{black}
\begin{tabular}{!{\color{black}\vrule}l!{\color{black}\vrule}l!{\color{black}\vrule}l!{\color{black}\vrule}l!{\color{black}\vrule}l!{\color{black}\vrule}l!{\color{black}\vrule}l!{\color{black}\vrule}}
\hline
 \textit{$v_x$}  &  \textit{$v_y$}  &    \textit{$G_S$}    &    \textit{$G_W$}    &    \textit{$G_R$}    &    \textit{$G_C$}    &  \textit{$G$}   \\
\hline
Faye Wu       & Fei Wu        & 1.0000 & 1.0000 & 1.0000 & 1.0000 & 1.0000        \\
\hline
\end{tabular}
\arrayrulecolor{black}
\label{table_16}
\end{table}

\subsubsection{Redundant vertices merging}
After vertices screening we merged redundant vertices $v_\text{Faye WuV}$ and $v_\text{Fei Wu}$ and their relations, then we got academic network $G'$ without redundant information. In Figure, we can find that $v_\text{Fei Wu}$ and its relations were removed by vertices merging, but $v_\text{Faye Wu}$ is saved. Likewise, we can save $v_\text{Faye Wu}$ and remove $v_\text{Fei Wu}$ in the process. Compared to the network before merging, the relations between $v_\text{Faye Wu}$ and neighbors were not changed. It indicated that the vertices merging was correct.

The analyzing above indicates that we can promote accuracy of vertices uniqueness identifying based on structure error calculation and transaction activity paths similarity. Similarity threshold setting implements vertices screening, which is the basis of redundant vertices merging. Therefore, our solution realized character correction in social networks from multimedia datasets.

\section{Conclusion}
\label{Conclusion}

In this paper, we introduce the framework of social network modeling via multimedia data. Then, we present the notion of structure error according to structure features of networks and vertices merging principles, and then calculated structure error and screened out redundant vertices by using transaction information to build social networks. Besides, we designed algorithm of vertices similarity which can precisely measure character vertices uniqueness and created redundant vertices set. Finally, we removed redundant information in a network by merging redundant vertices in the set. Our solution improved the accuracy of character uniqueness recognition, and solved character correction effectively in a network. At present, we set threshold empirical during experiment, but we do not implement intellectualized adaptive adjustment of threshold yet. In future work, we will compute the range of threshold based upon large amount of network data and statistical techniques, and then design adaptive adjustment algorithm.

\textbf{Acknowledgments:} This work was supported in part by the National Natural Science Foundation of China (61272150, 61379110, 61472450, 61402165, 61702560, S1651002, M1450004), the Key Research Program of Hunan Province(2016JC2018), and Science and Technology Plan of Hunan Province Project (201
8JJ2099, 2018JJ3691)



\bibliographystyle{spmpsci}      

\bibliography{ref}

\begin{thebibliography}{10}
\providecommand{\url}[1]{{#1}}
\providecommand{\urlprefix}{URL }
\expandafter\ifx\csname urlstyle\endcsname\relax
  \providecommand{\doi}[1]{DOI~\discretionary{}{}{}#1}\else
  \providecommand{\doi}{DOI~\discretionary{}{}{}\begingroup
  \urlstyle{rm}\Url}\fi

\bibitem{DBLP:conf/www/BekkermanM05}
Bekkerman, R., McCallum, A.: Disambiguating web appearances of people in a
  social network.
\newblock In: Proceedings of the 14th international conference on World Wide
  Web, {WWW} 2005, Chiba, Japan, May 10-14, 2005, pp. 463--470 (2005)

\bibitem{DBLP:conf/globecom/DingZWG11}
Ding, X., Zhang, L., Wan, Z., Gu, M.: De-anonymizing dynamic social networks.
\newblock In: Proceedings of the Global Communications Conference, {GLOBECOM}
  2011, 5-9 December 2011, Houston, Texas, {USA}, pp. 1--6 (2011)

\bibitem{DBLP:SOSAmultimedia}
F, A., V, M., A, P.: Sos: A multimedia recommender system for online social
  networks.
\newblock In: Future Generation Computer Systems (2017)

\bibitem{DBLP:journals/tmm/HuangLZZCZ17}
Huang, F., Li, X., Zhang, S., Zhang, J., Chen, J., Zhai, Z.: Overlapping
  community detection for multimedia social networks.
\newblock {IEEE} Trans. Multimedia \textbf{19}(8), 1881--1893 (2017)

\bibitem{DBLP:conf/ShanmeiTang}
Huang, F., Tang, S., Ling, C.X.: Extracting academic activity transaction in
  chinese documents.
\newblock In: Proceedings of the Eighth International Conference on Intelligent
  Systems and Knowledge Engineering, Shenzhen, China, (ISKE 2013), (2013)
  November, 278, pp125-135., year = {2013}

\bibitem{DBLP:conf/WenjieXiao}
Huang, F., Xiao, W., Zhang, H.: Visualization of clustered network graphs based
  on constrained optimization partition layout.
\newblock In: Proceedings of the Eighth International Conference on Intelligent
  Systems and Knowledge Engineering, Shenzhen, China, (ISKE 2013), (2013)
  November, 278, pp125-135., pages = {1381--1394}, year = {2014}

\bibitem{DBLP:conf/ic3/KatalWG13}
Katal, A., Wazid, M., Goudar, R.H.: Big data: Issues, challenges, tools and
  good practices.
\newblock In: Sixth International Conference on Contemporary Computing, {IC3}
  2013, Noida, India, August 8-10, 2013, pp. 404--409 (2013)

\bibitem{DBLP:conf/icwsm/KorayemC13}
Korayem, M., Crandall, D.J.: De-anonymizing users across heterogeneous social
  computing platforms.
\newblock In: Proceedings of the Seventh International Conference on Weblogs
  and Social Media, {ICWSM} 2013, Cambridge, Massachusetts, USA, July 8-11,
  2013. (2013)

\bibitem{DBLP:conf/mownet/LaforestSFCMLGS14}
Laforest, F., Sommer, N.L., Fr{\'{e}}not, S., de~Corbi{\`{e}}re, F.,
  Mah{\'{e}}o, Y., Launay, P., Gravier, C., Subercaze, J., Reimert, D., Brodu,
  E., Daikh, I., Phelippeau, N., Adam, X., Guidec, F., Grumbach, S.: {C3PO:}
  {A} spontaneous and ephemeral social networking framework for a collaborative
  creation and publishing of multimedia contents.
\newblock In: International Conference on Selected Topics in Mobile and
  Wireless Networking, MoWNet 2014, Rome, Italy, September 8-9, 2014, pp.
  129--134 (2014)

\bibitem{DBLP:conf/Leicht}
Leicht, E., Holme, P., Newman, M.E.: Vertex similarity in networks.
\newblock In: Physical Review E 73, no. 2 (2006): 026120 (2006)

\bibitem{DBLP:conf/sigmod/LiuT08}
Liu, K., Terzi, E.: Towards identity anonymization on graphs.
\newblock In: Proceedings of the {ACM} {SIGMOD} International Conference on
  Management of Data, {SIGMOD} 2008, Vancouver, BC, Canada, June 10-12, 2008,
  pp. 93--106 (2008)

\bibitem{DBLP:conf/sp/NarayananS08}
Narayanan, A., Shmatikov, V.: Robust de-anonymization of large sparse datasets.
\newblock In: 2008 {IEEE} Symposium on Security and Privacy (S{\&}P 2008),
  18-21 May 2008, Oakland, California, {USA}, pp. 111--125 (2008)

\bibitem{DBLP:conf/sp/NarayananS09}
Narayanan, A., Shmatikov, V.: De-anonymizing social networks.
\newblock In: 30th {IEEE} Symposium on Security and Privacy (S{\&}P 2009),
  17-20 May 2009, Oakland, California, {USA}, pp. 173--187 (2009)

\bibitem{DBLP:journals/jvcir/NiePWZS17}
Nie, W., Peng, W., Wang, X., Zhao, Y., Su, Y.: Multimedia venue semantic
  modeling based on multimodal data.
\newblock J. Visual Communication and Image Representation \textbf{48},
  375--385 (2017)

\bibitem{DBLP:journals/zhaosgaoy}
S, Z., Y, G., G, D.: Real-time multimedia social event detection in microblog.
\newblock IEEE Transactions on Cybernetics  (2017)

\bibitem{DBLP:conf/cts/SagirogluS13}
Sagiroglu, S., Sinanc, D.: Big data: {A} review.
\newblock In: 2013 International Conference on Collaboration Technologies and
  Systems, {CTS} 2013, San Diego, CA, USA, May 20-24, 2013, pp. 42--47 (2013)

\bibitem{DBLP:conf/bigmm/SangX15}
Sang, J., Xu, C.: On analyzing the 'variety' of big social multimedia.
\newblock In: 2015 {IEEE} International Conference on Multimedia Big Data,
  BigMM 2015, Beijing, China, April 20-22, 2015, pp. 5--8 (2015)

\bibitem{DBLP:conf/ism/SangXJ16}
Sang, J., Xu, C., Jain, R.: Social multimedia ming: From special to general.
\newblock In: {IEEE} International Symposium on Multimedia, {ISM} 2016, San
  Jose, CA, USA, December 11-13, 2016, pp. 481--485 (2016)

\bibitem{DBLP:conf/ccs/SrivatsaH12}
Srivatsa, M., Hicks, M.: Deanonymizing mobility traces: using social network as
  a side-channel.
\newblock In: the {ACM} Conference on Computer and Communications Security,
  CCS'12, Raleigh, NC, USA, October 16-18, 2012, pp. 628--637 (2012)

\bibitem{DBLP:journals/pvldb/SunHYYW11}
Sun, Y., Han, J., Yan, X., Yu, P.S., Wu, T.: Pathsim: Meta path-based top-k
  similarity search in heterogeneous information networks.
\newblock {PVLDB} \textbf{4}(11), 992--1003 (2011)

\bibitem{DBLP:conf/kdd/TangZYLZS08}
Tang, J., Zhang, J., Yao, L., Li, J., Zhang, L., Su, Z.: Arnetminer: extraction
  and mining of academic social networks.
\newblock In: Proceedings of the 14th {ACM} {SIGKDD} International Conference
  on Knowledge Discovery and Data Mining, Las Vegas, Nevada, USA, August 24-27,
  2008, pp. 990--998 (2008)

\bibitem{YXWINF13}
Wang, Y., Huang, X., Wu, L.: Clustering via geometric median shift over
  riemannian manifolds.
\newblock Information Sciences \textbf{220}, 292--305 (2013)

\bibitem{DBLP:conf/mm/WangLWZ15}
Wang, Y., Lin, X., Wu, L., Zhang, W.: Effective multi-query expansions: Robust
  landmark retrieval.
\newblock In: Proceedings of the 23rd Annual {ACM} Conference on Multimedia
  Conference, {MM} '15, Brisbane, Australia, October 26 - 30, 2015, pp. 79--88
  (2015)

\bibitem{DBLP:journals/tip/WangLWZ17}
Wang, Y., Lin, X., Wu, L., Zhang, W.: Effective multi-query expansions:
  Collaborative deep networks for robust landmark retrieval.
\newblock {IEEE} Trans. Image Processing \textbf{26}(3), 1393--1404 (2017)

\bibitem{DBLP:conf/mm/WangLWZZ14}
Wang, Y., Lin, X., Wu, L., Zhang, W., Zhang, Q.: Exploiting correlation
  consensus: Towards subspace clustering for multi-modal data.
\newblock In: Proceedings of the {ACM} International Conference on Multimedia,
  {MM} '14, Orlando, FL, USA, November 03 - 07, 2014, pp. 981--984 (2014)

\bibitem{DBLP:conf/sigir/WangLWZZ15}
Wang, Y., Lin, X., Wu, L., Zhang, W., Zhang, Q.: {LBMCH:} learning bridging
  mapping for cross-modal hashing.
\newblock In: Proceedings of the 38th International {ACM} {SIGIR} Conference on
  Research and Development in Information Retrieval, Santiago, Chile, August
  9-13, 2015, pp. 999--1002 (2015)

\bibitem{DBLP:journals/tip/WangLWZZH15}
Wang, Y., Lin, X., Wu, L., Zhang, W., Zhang, Q., Huang, X.: Robust subspace
  clustering for multi-view data by exploiting correlation consensus.
\newblock {IEEE} Trans. Image Processing \textbf{24}(11), 3939--3949 (2015)

\bibitem{DBLP:conf/cikm/WangLZ13}
Wang, Y., Lin, X., Zhang, Q.: Towards metric fusion on multi-view data: a
  cross-view based graph random walk approach.
\newblock In: 22nd {ACM} International Conference on Information and Knowledge
  Management, CIKM'13, San Francisco, CA, USA, October 27 - November 1, 2013,
  pp. 805--810 (2013)

\bibitem{DBLP:conf/pakdd/WangLZW14}
Wang, Y., Lin, X., Zhang, Q., Wu, L.: Shifting hypergraphs by probabilistic
  voting.
\newblock In: Advances in Knowledge Discovery and Data Mining - 18th
  Pacific-Asia Conference, {PAKDD} 2014, Tainan, Taiwan, May 13-16, 2014.
  Proceedings, Part {II}, pp. 234--246 (2014)

\bibitem{YXJPAKDD14}
Wang, Y., Pei, J., Lin, X., Zhang, Q., Zhang, W.: An iterative fusion approach
  to graph-based semi-supervised learning from multiple views.
\newblock In: PAKDD (2014)

\bibitem{DBLP:journals/corr/abs-1708-02288}
Wang, Y., Wu, L.: Beyond low-rank representations: Orthogonal clustering basis
  reconstruction with optimized graph structure for multi-view spectral
  clustering.
\newblock Neural Networks \textbf{103}, 1--8 (2018)

\bibitem{NNLS2018}
Wang, Y., Wu, L., Lin, X., Gao, J.: Multiview spectral clustering via
  structured low-rank matrix factorization.
\newblock {IEEE} Trans. Neural Networks and Learning Systems  (2018)

\bibitem{DBLP:conf/ijcai/WangZWLFP16}
Wang, Y., Zhang, W., Wu, L., Lin, X., Fang, M., Pan, S.: Iterative views
  agreement: An iterative low-rank based structured optimization method to
  multi-view spectral clustering.
\newblock In: Proceedings of the Twenty-Fifth International Joint Conference on
  Artificial Intelligence, {IJCAI} 2016, New York, NY, USA, 9-15 July 2016, pp.
  2153--2159 (2016)

\bibitem{DBLP:journals/tnn/WangZWLZ17}
Wang, Y., Zhang, W., Wu, L., Lin, X., Zhao, X.: Unsupervised metric fusion over
  multiview data by graph random walk-based cross-view diffusion.
\newblock {IEEE} Trans. Neural Netw. Learning Syst. \textbf{28}(1), 57--70
  (2017)

\bibitem{DBLP:journals/mta/WuHZSW15}
Wu, L., Huang, X., Zhang, C., Shepherd, J., Wang, Y.: An efficient framework of
  bregman divergence optimization for co-ranking images and tags in a
  heterogeneous network.
\newblock Multimedia Tools Appl. \textbf{74}(15), 5635--5660 (2015)

\bibitem{DBLP:journals/ivc/WuW17}
Wu, L., Wang, Y.: Robust hashing for multi-view data: Jointly learning low-rank
  kernelized similarity consensus and hash functions.
\newblock Image Vision Comput. \textbf{57}, 58--66 (2017)

\bibitem{DBLP:journals/pr/WuWGL18}
Wu, L., Wang, Y., Gao, J., Li, X.: Deep adaptive feature embedding with local
  sample distributions for person re-identification.
\newblock Pattern Recognition \textbf{73}, 275--288 (2018)

\bibitem{DBLP:journals/cviu/WuWGHL18}
Wu, L., Wang, Y., Ge, Z., Hu, Q., Li, X.: Structured deep hashing with
  convolutional neural networks for fast person re-identification.
\newblock Computer Vision and Image Understanding \textbf{167}, 63--73 (2018)

\bibitem{TC2018}
Wu, L., Wang, Y., Li, X., Gao, J.: Deep attention-based spatially recursive
  networks for fine-grained visual recognition.
\newblock {IEEE} Trans. Cybernetics  (2018)

\bibitem{DBLP:journals/pr/WuWLG18}
Wu, L., Wang, Y., Li, X., Gao, J.: What-and-where to match: Deep spatially
  multiplicative integration networks for person re-identification.
\newblock Pattern Recognition \textbf{76}, 727--738 (2018)

\bibitem{LYSARX}
Wu, L., Wang, Y., Shao, L.: Cycle-consistent deep generative hashing for
  cross-modal retrieval.
\newblock In: arXiv:1804.11013 (2018)

\bibitem{LYJMM13}
Wu, L., Wang, Y., Shepherd, J.: Efficient image and tag co-ranking: a bregman
  divergence optimization method.
\newblock In: ACM Multimedia, pp. 593--596 (2013)

\bibitem{DBLP:journals/tkde/WuZW014}
Wu, X., Zhu, X., Wu, G., Ding, W.: Data mining with big data.
\newblock {IEEE} Trans. Knowl. Data Eng. \textbf{26}(1), 97--107 (2014)

\bibitem{DBLP:journals/vlc/YeXDWLZ14}
Ye, C., Xiong, Z., Ding, Y., Wang, G., Li, J., Zhang, K.: Joint fingerprinting
  and encryption in hybrid domains for multimedia sharing in social networks.
\newblock J. Vis. Lang. Comput. \textbf{25}(6), 658--666 (2014)

\bibitem{DBLP:journals/chb/ZhuhadarYL13}
Zhuhadar, L., Yang, R., Lytras, M.D.: The impact of social multimedia systems
  on cyberlearners.
\newblock Computers in Human Behavior \textbf{29}(2), 378--385 (2013)

\end{thebibliography}

\end{document}